\documentclass[6pt, preprint2]{aastex}
\usepackage[]{natbib}
\begin{document}
\title{Discrete Components in the Radial Velocities of ScI Galaxies}

\author{M.B. Bell\altaffilmark{1}, S.P. Comeau\altaffilmark{1} and D.G. Russell\altaffilmark{2}}

\altaffiltext{1}{Herzberg Institute of Astrophysics,
National Research Council of Canada, 100 Sussex Drive, Ottawa,
ON, Canada K1A 0R6}
\altaffiltext{2}{Owego Free Academy, Owego, NY 13827, USA}

\begin{abstract}

We examine the V$_{\rm CMB}$ velocities and Tully-Fisher distances of 83 ScI galaxies and find that their velocities contain the same discrete velocity components identified previously by Tifft. After removing the discrete velocity components, the scatter remaining on the Hubble plot is in good agreement with the Tully-Fisher distance uncertainties. Although there is, as yet, no physical explanation for these discrete components, the fact that they appear to be identical to those found previously by Tifft suggests that their reality needs to be taken seriously.

\end{abstract}



\section{Introduction.}

\citet{tif76,tif77a,tif77b,tif96,tif97} found families of discrete velocities in normal galaxies and fitted them to a model in which they were all harmonically related to $c$ the speed of light. In his model the discrete components detected are octave-spaced sub-harmonics of $c$, and related harmonics, and the redshift is assumed to arise from a time-dependent decay from an origin at the Planck scale. His observed velocity periods can be accurately described by the relation P = $c2^{(-9D+T)/9}$, where D is the number of doublings and T defines the different families. There can be many different families but the most common, or basic, of these make up what is referred to by \citet{tif96,tif97} as the T = 0 family of discrete values. These were found in common spiral galaxies. They showed discrete velocity components near 4.5, 9, 18, 36, 72, 145, 290, etc., km s$^{-1}$ (hereafter Tifft's T=0 period). Although the other period groups were less common, they showed this same octave-related, or doubling nature.

We have recently examined the radial velocities and distances of 55 Spiral galaxies and 36 Type Ia Supernovae (SneIa) galaxies with accurate distances \citep{bel03a}. Both groups showed evidence for discrete, intrinsic velocity components related to those reported by Tifft. However, the galaxies we studied extended to much larger distances and showed that the entire radial velocity was not quantized, as previously thought by Tifft. Instead, the discrete components \em were superimposed on top of the Hubble flow. \em The observed discrete spectral shifts were always positive (redshifts). The origin of these discrete velocities is unknown. When the intrinsic components were removed from the measured radial velocities, a Hubble constant of H$_{\rm o}$ = 58 km s$^{-1}$ Mpc$^{-1}$ was obtained for the local Universe, and the velocity-distance plot contained less scatter. This H$_{o}$-value is 20\% lower than that obtained in the Hubble Key Project \citep{fre01}. The difference between these two Hubble constants is due to the fact that \citet{fre01} interpreted the scatter on the Hubble plot as random peculiar velocities containing both positive and negative Doppler components. 

Because what is meant by "peculiar velocity" is of fundamental importance to this analysis we define the following two terms to avoid future confusion. The term "peculiar velocities" is used here in the conventional sense to refer to the velocity scatter about the Hubble line and will hereafter be referred to as PVs. Because we assume here that these PVs may contain discrete intrinsic components that can cause them to fall along lines that run parallel to the Hubble line, our analysis involves the determination of the RMS deviation relative to the nearest of these intrinsic lines instead of one Hubble line. We define the scatter about these discrete lines as the "true peculiar velocities" and they are hereafter referred to as TPVs

Here we examine the radial velocities of 83 ScI galaxies. We use the same analysis procedure as used previously so the results can be easily compared. We first describe the procedure in detail in order to convince the reader that we are really detecting discrete velocity components.

\section{The Data}

B- and I-band Tully-Fisher (TF) distances, and V$_{\rm CMB}$ velocities for the 83 ScI galaxies were obtained from Table VI of \citet{rus03}. The TF distances for these galaxies were calculated utilizing the Type dependent TF relations of \citet{rus03}, who found that galaxies of morphological types SbcI, Sbc I-II, ScI, ScI-II and Seyferts have systematically underestimated TF distances utilizing a standard single TF equation.  The 83 ScI galaxies were drawn from the samples of \citet{mat96} and \citet{gio97} and are all classified as Sbc or Sc of luminosity class I or I-II in the LEDA database.  TF scatter was significantly reduced with the use of type dependent TF relations for the calibrator samples \citep{rus03} and for cluster samples \citep{rus04}. The TF scatter of the calibrators in the B-band was only $\pm0.09$ mag for the ScI group and $\pm0.16$ mag for the Sb/ScIII group when using the type dependent TF relations compared with $\pm0.30$ mag with a traditional single calibration \citep{rus03}. 
The scatter quoted here represents the empirical scatter of the TF relation relative to calibrator.


\section{Analysis}

From our previous work and Tifft's, we can make the following initial assumptions to limit the number of free parameters. First, The Hubble slope obtained must be H$_{\rm o}$ = 58 km s$^{-1}$ Mpc$^{-1}$. Second, since we are examining spiral galaxies, the discrete components present will be predominantly those predicted by Tifft's most common (T = 0) group, and third, the discrete components will be superimposed on top of the Hubble flow. These assumptions severely restrict what the data must show if our previous results are to be confirmed. The first of these assumptions is an important one because it allows us to differentiate between real and random features.

As in previous analyses \citep{bel03a}, in a given source sample, V$_{\rm CMB}$, the radial velocity of each source relative to the cosmic microwave background reference frame, was first plotted vs distance. Fig 1 shows such a plot using peculiar velocities (PVs) that were randomly generated. The Hubble slope used in the generation of this sample (H$_{\rm o}$ = 85) was taken from the actual source distribution and is the slope obtained for the data when the lines in Fig 1 are ignored and a linear relation is fitted to the data points. In this figure the solid lines represent Tifft's relevant T = 0 discrete components superimposed on top of an H$_{\rm o}$ = 58 Hubble slope. Since the superimposed velocities were randomly generated there should be no evidence in Fig 1 that the source velocities fall along the discrete velocity grid lines, and clearly there is not. As in all previous analyses we then calculated the RMS deviation of the sources from their nearest discrete velocity grid line. We then repeated this calculation for a range of Hubble constants. When the velocity distribution does not contain Tifft's discrete components, then, as the Hubble slope is varied, \em one \em RMS minimum will be obtained. \em This minimum will occur for that Hubble slope that positions the densest portion of the discrete velocity grid lines on top of the densest portion of the source distribution. \em The resulting RMS deviation plot is shown in Fig 2. As predicted, one RMS minimum is obtained and it occurs near a Hubble slope of H$_{\rm o}$ = 70. This then, is how the RMS deviation plot is expected to appear when Tifft's discrete velocity components are $not$ present in the radial velocities.


\begin{figure}
\hspace{-1.0cm}
\vspace{-2.0cm}
\epsscale{1.0}
\plotone{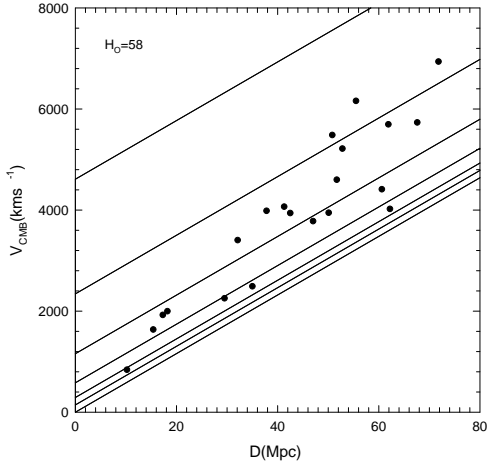}
\caption{Test set of randomly generated V$_{\rm CMB}$ velocities plotted vs distance. Solid lines represent Tifft's T=0 discrete velocities for H$_{\rm o}$ = 58. \label{fig1}}
\end{figure}


\begin{figure}
\hspace{-1.0cm}
\vspace{-2.0cm}
\epsscale{1.0}
\plotone{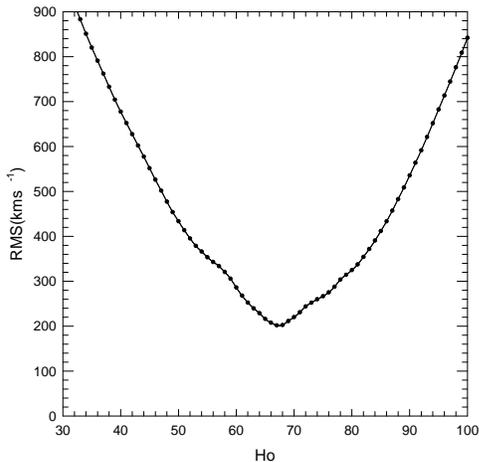}
\caption{RMS deviation in test V$_{\rm CMB}$ velocities about Tifft's T=0 discrete velocity lines for data in Fig 1 plotted as a function of H$_{\rm o}$. Only one minimum is visible in the plot. \label{fig2}}
\end{figure}


We now demonstrate how the RMS deviation plot in Fig 2 will be affected if Tifft's T = 0 discrete velocities $are$ present in the source velocities. To do this we generated a second data set containing Tifft's T = 0 components superimposed on top of the Hubble slope. The generated velocities contained no distance errors, no true peculiar velocities (TPVs), and were superimposed on top of a Hubble slope of H$_{\rm o}$ = 50. A slope of 50 was used to show that there is nothing magical about the value 58. Fig 3 shows these velocities plotted versus distance, using the same distance values as in Fig 1. In Fig 3 the velocities are clearly aligned along Tifft's T = 0 discrete velocity lines for H$_{\rm o}$ = 50. The RMS deviation of the source velocities from their nearest discrete velocity lines in Fig 3 was then calculated for a range of Hubble slopes as was done in Fig 2. As the Hubble slope is varied, a low RMS value will be obtained as before for that H$_{\rm o}$-value that positions the densest portion of the lines near the center of the source velocity distribution. This is again expected to occur near H$_{\rm o}$ = 70. However, with Tifft's  discrete components present in the data, a SECOND dip in the RMS deviation plot will occur at the Hubble slope where the source velocities are positioned along the discrete velocity grid lines. The resulting RMS deviation plot is shown in Fig 4, where a second dip is now visible corresponding to a Hubble slope of 50, indicated by the arrow.

This SECOND dip in the RMS deviation curve at H$_{\rm o}$ = 50 in Fig 4 is thus THE DEFINING SIGNATURE OF DISCRETE VELOCITIES IN THE DATA. The scatter in the source distances about the discrete velocity lines will determine the depth of the second RMS dip. This RMS dip therefore does not have to be deeper than the first dip near H$_{\rm o}$ = 70 to indicate that discrete velocities are present. For real data, however, it does have to occur at, or very near, H$_{\rm o}$ = 58, since this is the value found in all previous analyses of real galaxy data. Although it is theoretically possible for an RMS dip to be produced by random fluctuations in the source distances and peculiar velocities, if these fluctuations are truly random, \em so will be the value of the Hubble constant at which the RMS dip occurs. \em  Also, the amplitude of random dips is expected to decrease, as the number of sources in the sample increases.

\section{Tully-Fisher Distance Uncertainties}

Since, as noted above, the depth of the discrete component signature (i.e. the second RMS dip) is directly proportional to the size of the distance uncertainties, as these increase the second dip will become shallower, until it eventually disappears, as in Fig 2. There are good reasons to suspect that some of the sources will have larger distance errors than others. As long as these sources are present in the sample they are expected to reduce the visibility of the second RMS dip. We therefore first looked at a test sample drawn from the 83 ScI galaxies that we have a good reason to believe will have some of the most accurate distances.


\begin{figure}
\hspace{-1.0cm}
\vspace{-2.0cm}
\epsscale{1.0}
\plotone{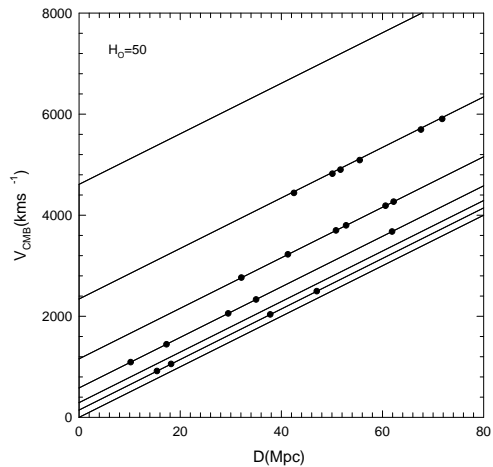}
\caption{Test set of V$_{\rm CMB}$ velocities containing only Tifft's T = 0 discrete components, plotted vs distance. Solid lines represent Tifft's T=0 discrete velocities. \label{fig3}}
\end{figure}


\begin{figure}
\hspace{-1.0cm}
\vspace{-2.0cm}
\epsscale{1.0}
\plotone{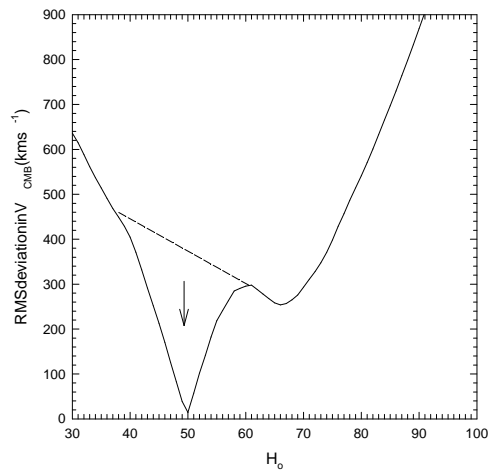}
\caption{RMS deviation in test V$_{\rm CMB}$ velocities about Tifft's T=0 discrete velocity lines for the velocity distribution in Fig 3, plotted versus H$_{\rm o}$. Two RMS dips are now visible in the plot, one at H$_{\rm o}$ = 50 and one near H$_{\rm o}$ = 70. The arrow indicates the location of the dip produced by the presence of Tifft's discrete velocities. \label{fig4}}
\end{figure}


By taking the difference between the B- and I-band distances for each source it is found that these differences range from 0 to 57.6 Mpc for the 83 ScI galaxies. Clearly, from this result, some distances have to have large errors. By making the simple assumption that the distances are likely to be more accurate when both B and I-Band distances agree, we obtained a sample that we felt should have accurate distances by choosing the 23 sources with the smallest B-I differences. The mean B-I difference for this sample was 0.8 Mpc.

In Fig 5 the V$_{\rm CMB}$ velocities of these 23 sources are plotted vs their B-band distances. Tifft's T = 0 discrete velocity lines are included on the plot. A solid line joins sources assumed to contain the same discrete velocity component. It is apparent from this plot that the sources are aligned along the lines representing the discrete velocity components for a Hubble slope near 58. If most of the scatter in the plot is then due to the discrete intrinsic components, the real dispersion (due to TPVs and distance errors) will be given by the scatter about the discrete lines. It must be much smaller and will not be clearly visible until the discrete components are removed (see Fig 7). The measurement error for velocity is likely to be less than 35 km s$^{-1}$ \citep{jor95}. In fact, the mean velocity error in Fig 5 is 17.6 km s$^{-1}$ from the velocity errors listed in the LEDA database for these sources. Although magnitude measurement errors are also known, they are expected to bear little resemblance to the true distance errors that are obtained after the various corrections are applied.

The RMS deviation plot for the sources in Fig 5 is given by the solid curve in Fig 6 and it also clearly shows the signature of Tifft's discrete velocities by the presence of a second RMS dip near H$_{\rm o}$ = 58. Also plotted in Fig 6 are two other curves. The dashed curve is the RMS deviation plot obtained by replacing the PVs of the 23 sources in Fig 5 with randomly generated ones. The dotted curve is an RMS deviation plot obtained by retaining the original PVs and generating random distance errors. Both random data sets were generated so that the resulting scatter was similar to that for the real data (assuming one Hubble line with no discrete components). Both plots obtained using randomly generated data show only one minimum. No second minimum is expected since Tifft's T = 0 components are not present in the randomly generated data.


\begin{figure}
\hspace{-1.0cm}
\vspace{-2.5cm}
\epsscale{1.0}
\plotone{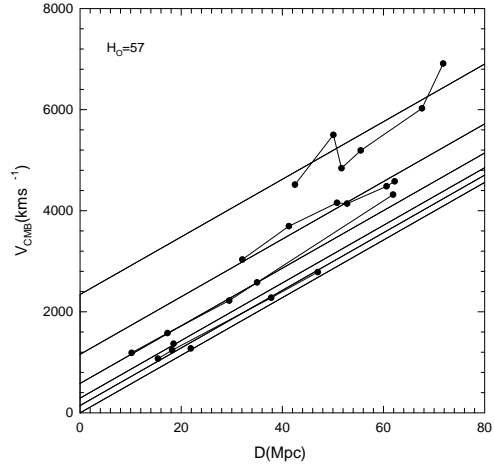}
\caption{Plot of V$_{\rm CMB}$ velocities as a function of distance for the 23 ScI galaxies with smallest B-I distance differences from Table VI of \citet{rus03}. \label{fig5}}
\end{figure}


\begin{figure}
\hspace{-1.0cm}
\vspace{-2.5cm}
\epsscale{1.0}
\plotone{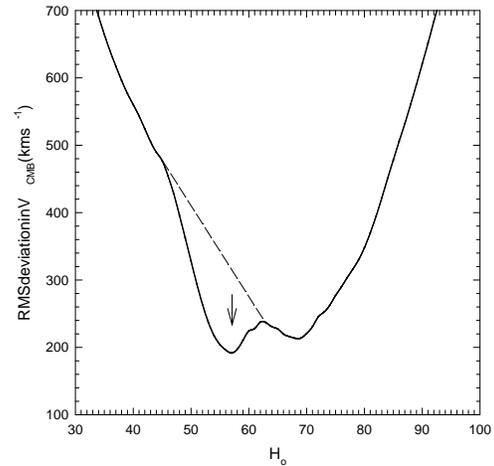}
\caption{(solid curve) Plot of the RMS deviation in V$_{\rm CMB}$ about the discrete velocity lines vs H$_{\rm o}$ for 23 ScI galaxies in Fig 5, showing a second RMS dip at H$_{\rm o}$ = 58. (dashed curve) Similar plot for sources with randomly generated peculiar velocities. (dotted curve) Similar plot for sources with randomly generated distance errors as described in the text. \label{fig6}}
\end{figure}


In Fig 7 the discrete components (as defined by the lines in Fig 5) have been removed from the radial velocities leaving only the component V$_{\rm H}$, due to the Hubble flow, with its uncertainty in distance and any small TPVs remaining. The dashed lines in Fig 7 are assumed to represent the peak-to-peak scatter in the distances, as a function of distance, for the 23 galaxies with smallest B-I differences. The peak-to-peak scatter in Fig 7 corresponds to a 1 $\sigma$ uncertainty in the distance modulus of $\pm0.2$ mag, if H$_{\rm o}$ = 58, assuming no TPVs. After conversion to distance, these uncertainties (1 sigma) have been included in Fig 7. This is consistent with the Cepheid calibrators for which the 12 ScI group calibrators had a B-band scatter of $\pm0.09$ mag (1 sigma). The 23 galaxies in Fig 7 represent those galaxies in the sample that are likely to have some of the smallest distance uncertainties. The scatter obtained for the entire sample of 83 galaxies is expected to be larger. Typical scatter, when type dependence is not accounted for is at least $\pm0.25$ magnitude \citep{rus03}. When type dependence is taken into account, as here, our value of $\pm0.2$ magnitude is in good agreement with what is expected. These results, after removal of the discrete velocity components, indicate that any TPV components present cannot be large.

The solid line in Fig 7 represents a linear fit to the data which gave a Hubble constant of H$_{\rm o} = 57\pm2$.
Previous analyses of galaxies gave values of H$_{\rm o}$ = 57.9 for SneIa galaxies, 57.5 for Sb galaxies and 60.0 for Sc galaxies  \citep{bel03a}. With equal weighting these results give a mean value of H$_{\rm o}$ = 58.1$ \pm$1.2 km s$^{\rm -1}$ Mpc$^{\rm -1}$.

\section{RMS Deviation Plot for all 83 ScI Galaxies}

Fig 8 shows the plots obtained for the RMS deviation in V$_{\rm CMB}$ velocities for all 83 ScI galaxies for both B- and I-band TF distances. There is no evidence in the I-Band curve for a second RMS deviation dip at H$_{\rm o}$ = 58, and the minimum RMS deviation found is near 580 km s$^{-1}$. In the B-Band curve there is a clear second dip near H$_{\rm o}$ = 58, indicated by the arrow, and its minimum RMS deviation is now close to 420 km s$^{-1}$. Although this is much higher than the RMS-value of 190 obtained for the 23 sources selected because their B and I-Band distances were similar, this is not surprising since the full sample contains sources with a wide range of inclination angles and rotation velocities. It is well known that distance errors can vary significantly with inclination \citep{sak00,tul00}.



\begin{figure}
\hspace{-1.0cm}
\vspace{-2.0cm}
\epsscale{1.0}
\plotone{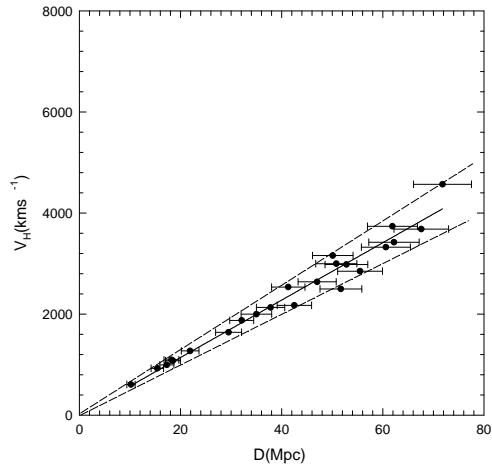}
\caption{Plot of the residual velocity V$_{\rm H}$ vs B-band distance after removal of the discrete velocity components in Fig 5. \label{fig7}}
\end{figure}


\begin{figure}
\hspace{-1.0cm}
\vspace{-2.0cm}
\epsscale{1.0}
\plotone{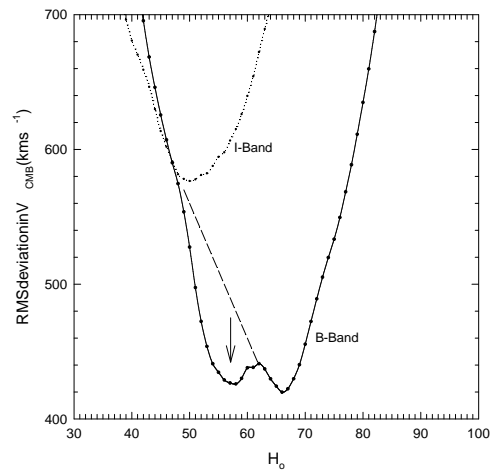}
\caption{RMS deviation in V$_{\rm CMB}$ about the discrete velocity lines for all 83 ScI galaxies when (dotted curve) using I-Band distances and, (solid curve) using B-Band distances. \label{fig8}}
\end{figure}


\section{Discussion}

One of the strongest arguments favoring our interpretation that the detection of a second dip in the RMS deviation in V$_{\rm CMB}$ as H$_{\rm o}$ is varied is due to discrete velocities present in the data, is the fact that in the present analysis, as well as in all previous analyses where galaxy TF distances were examined, this second dip has always occurred at the same Hubble slope of H$_{\rm o}$ = 58. Although the occurrence of RMS dips due to random clumping in the source distribution is possible, there are few arguments that can be made to explain the consistent appearance of randomly produced dips at the same Hubble slope. There would appear to be only two reasons why dips would always appear at the same Hubble slope in the real data. One is if the discrete components are real, and the other is if the analysis procedure is somehow affecting the result. The latter case was examined extensively in Test 2 of a previous paper \citep{bel03b}, where it was clearly demonstrated, by averaging RMS deviation curves for 10 randomly generated samples, that our reduction procedure was not introducing a second RMS dip at a particular Hubble slope. In Test 1 of that paper we also demonstrated that the randomly generated samples gave only one RMS dip. We note again that the depth of the second RMS dip, the signature of the presence of discrete components, is unrelated to the overall scatter in the data (i.e. to the size of the PVs generated). It is entirely dependent upon how closely aligned the data points are to Tifft's discrete velocity lines (i.e. how small the TPVs and distance errors are). If a second RMS dip appears in the real data it means that the PVs cannot be purely random, but must contain discrete components that align the sources along the "discrete velocity" lines.  We have further demonstrated again in Fig 6 of this paper that the second RMS dip is not seen when the discrete components are not present in the data, and it is therefore not produced by the analysis procedure. We have also demonstrated in Figs 3 and 4 that the location of the second dip is determined by the data. We conclude that the second RMS dip at H$_{\rm o}$ = 58 that we have now detected in four completely separate galaxy groups is due to the presence of discrete velocity components in the radial velocities of these galaxies. The fact that the discrete velocities found are identical to those reported previously by Tifft strongly reinforces this claim.


\section{Conclusions}

We have examined the velocities and distances of a source sample containing 83 ScI galaxies. We find evidence that the velocity components of Tifft's most common T=0 velocity period are present in the radial velocities of these galaxies. We find for a particular sub-sample of 23 ScI galaxies, chosen because their distances are likely to be accurate, that after the discrete velocities are removed, the distance scatter remaining in the Hubble plot corresponds to a mean magnitude error in their distance modulus of $\pm0.20$ magnitude. This is in good agreement with the value expected when morphological type is taken into consideration. We also find, for the 83 sources, that B-band distances are considerably more accurate than I-band. As a result of this we suggest that our technique may be a valuable one to use to study the Tully-Fisher distance errors as a function of galaxy inclination and rotation velocity.

We thank Dr. D. McDiarmid for helpful suggestions.


\end{document}